# Imaging Topological Defect Dynamics Mediating 2D Skyrmion Lattice Melting

Raphael Gruber[1], Jan Rothörl[1], Simon M. Fröhlich[1], Maarten A. Brems[1], Fabian Kammerbauer[1], Maria-Andromachi Syskaki[1,2], Elizabeth M. Jefremovas[1], Sachin Krishnia[1], Asle Sudbø[3], Peter Virnau[1], Mathias Kläui[1,3]*

1. Institute of Physics, Johannes Gutenberg-Universität Mainz, Staudingerweg 7, 55128 Mainz, Germany.
2. Singulus Technologies AG, Hanauer Landstraße 103, 63796 Kahl am Main, Germany.
3. Center for Quantum Spintronics, Department of Physics, Norwegian University of Science and Technology, 7491 Trondheim, Norway.

*Email: klaeui@uni-mainz.de

## Abstract

Topological defects are the key feature mediating 2D phase transitions. However, both resolution and tunability have been lacking to access the dynamics of the transitions. With dynamic Kerr microscopy, we directly capture the melting of a confined 2D magnetic skyrmion lattice with high resolution in real-time and -space. Skyrmions in magnetic thin films are two-dimensional, topologically non-trivial quasi-particles that provide rich dynamics as well as unique tunability as an essential ingredient for controlling phase behavior: We tune the skyrmion size and effective temperature on the fly to drive the two-step melting through an intermediate hexatic regime between the solid lattice and the isotropic liquid. We quantify the characteristic occurrence of topological defects mediating the transitions and reveal the so-far inaccessible dynamics of the lattice dislocations. The full real-time and -space imaging reveals the diffusion coefficient of the dislocations, which we find to be orders of magnitudes higher than that of the skyrmions.

# Introduction

Magnetic skyrmions are topologically non-trivial chiral spin structures exhibiting quasi-particle behavior[1-3]. Besides being ideal candidates for low-power applications in data storage and processing[4-13], skyrmions hosted in nanometer-thin films[3] are a good model system for studying 2D system properties[14-16].

Skyrmions in such thin film multilayer compounds are observable at room temperature by Kerr microscopy in both real-time and real-space[7,17,18], which is challenging in many other techniques[14,15]. Additionally, in the specially designed magnetic thin films, skyrmions exhibit thermally activated, Brownian-like diffusion[7]. Another key advantage of such systems is the unique tunability of the skyrmion size[17,19,20] and the skyrmion diffusivity[18] on the fly. With their purely repulsive interaction potentials[21-23], dense-packed skyrmions can also form ordered 2D lattices[14-16].

The Kosterlitz-Thouless-Halperin-Nelson-Young (KTHNY) theory[24-28] is an example describing 2D melting from a solid with translational quasi-long-range order (QLRO) to a disordered, isotropic liquid in two steps via an intermediate orientationally QLRO hexatic phase. The translational order is quantified by the translational correlation function

$$G_T(r = |\mathbf{r}_j - \mathbf{r}_k|) = \langle e^{-i\mathbf{G}\cdot(\mathbf{r}_j - \mathbf{r}_k)} \rangle$$

averaging the link between two particle positions $\mathbf{r}_j$ and $\mathbf{r}_k$ with respect to a reciprocal lattice vector $\mathbf{G}$ over the distance $r$. The orientational correlation function

$$G_6(r = |\mathbf{r}_j - \mathbf{r}_k|) = \langle \psi_6^*(\mathbf{r}_j)\psi_6(\mathbf{r}_k) \rangle$$

quantifies orientational order based on the local orientational order parameter

$$\psi_6(\mathbf{r}_j) = \frac{1}{N}\sum_{k=1}^{N} e^{-i6\theta_{jk}}$$

of a particle at position $\mathbf{r}_j$ with $N$ nearest neighbors labeled $k=1$ to $N$. $\theta_{jk}$ denotes the angle of the connecting vector $\mathbf{r}_k$-$\mathbf{r}_j$ with respect to an arbitrary axis[26].

In a 2D solid, $G_T(r)$ decays algebraically as $\propto r^{-\eta_T}$, signaling QLRO. When the exponent $\eta_T$ reaches its critical value of 1/3, an exponential decay $\propto \exp(-r/\xi_T)$ with the correlation length $\xi_T$ sets in; translational QLRO has disappeared[26]. In contrast, $G_6(r)$ is constant in a solid but shows an algebraic decay $\propto r^{-\eta_6}$ when translational order vanishes if orientational QLRO persists. Hence, orientational order is still present in what is referred to as the hexatic phase, which is unique to 2D systems[26,27]. When $\eta_6$ reaches its critical value of 1/4, also $G_6(r)$ becomes exponential $\propto \exp(-r/\xi_6)$ with a correlation length $\xi_6$. resulting in an isotropic liquid. At the transition from exponential to algebraic decay, the respective correlation lengths of both correlation functions diverge, causing the exponential term to vanish in the critical (QLRO) phases[26].

The two KTHNY phase transitions described above are associated with the unbinding and proliferation of pairs of topological defects of the lattice. Any lattice site with a number of nearest neighbors different from $N=6$ is a topological defect. A dislocation is a pair of defects with opposite topological charge: one $N=5$ and one $N=7$ defect. In a solid, only a few dislocation pairs occur, which are tightly bound and of opposite orientation. The orientation of a dislocation is specified by the Burger's vector. The Burger's vector is determined as the missing vector when encircling a dislocation counterclockwise with a set of lattice vectors, which would yield a closed path in a perfect lattice. At the transition point separating the solid from the hexatic phase, the dislocation pairs unbind and proliferate. This formation of isolated free dislocations causing the loss of translational QLRO is measurable macroscopically as a vanishing shear modulus $\mu$. At the transition point separating the hexatic from the liquid phase, the dislocations eventually unbind and proliferate into two isolated disclinations[25,26].

The existence of the three different phases as well as the detection of characteristic defects has been demonstrated in several systems and experiments, including colloids[29,30], superconducting vortices[31,32] and skyrmions[14,33,34]. However, in all these previous experimental investigations, resolution[14,15,31,32] or tunability[15,29,30,34] of the system have been insufficient to drive the system through phase transitions and elucidate the dynamics of the melting, including the defect evolution in real-time and -space.

For skyrmions in thin films, which exhibit rich dynamics and provide the required tunability[7,17,18], pinning effects in the form of a non-flat energy landscape in the magnetic multilayers have hampered the formation of QLRO of skyrmion lattices[15,16,33,35,36]. However, for sufficiently low pinning, the structural disorder may be small enough to allow for translational order while enhancing the phase space of the hexatic phase[33,35] enabling better feasibility to study the phases and transitions.

In this study, we exploit the unique on-the-fly tunability of our 2D skyrmion lattice to melt the system to disorder in a two-step process employing two independent methods: (i) by shrinking the skyrmions and reducing the packing fraction and (ii) by increasing the skyrmions' diffusivity corresponding to an effective temperature. We stabilize the skyrmion lattice in a low-pinning magnetic thin film stack[7,17,18] using a hexagonal geometric confinement providing commensurate boundary conditions[37] for the lattice formation as well as enabling rich dynamics. In combination with high-resolution imaging in real-time and -space, this yields powerful insights into the formation and dissociation of topological defects, which is the key feature mediating the melting transitions in 2D – but has so far been experimentally inaccessible.

# Skyrmion Lattice Melting in Two Steps

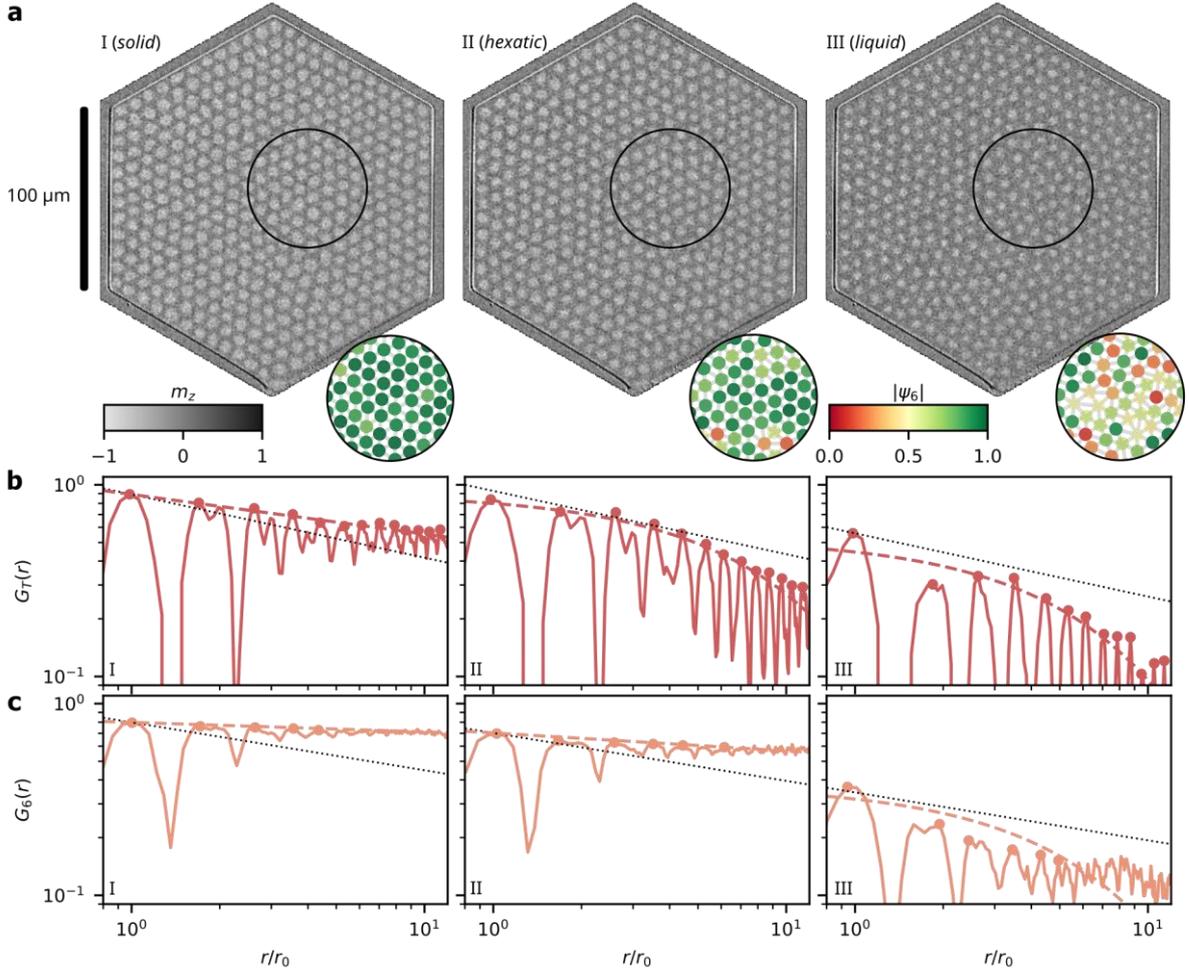

**Fig. 1. Skyrmion lattices with different order. a** Kerr microscopy images of 401 skyrmions comprising lattices in hexagonal geometric confinement: snapshots at (label I) 60 µT, (II) 84 µT and (III) 114 µT; gray-scale contrast represents the OOP magnetization $m_z$. The insets visualize the local order parameter $|\psi_6|$ per skyrmion in the black circle. **b** Translational correlation functions $G_T(r)$ as a function of the distance $r$ in units of nearest-neighbor distances $r_0$. The dashed line represents the power-law (I) and exponential (II-III) envelope fits, the dots are fitted points. The dotted black line is the power-law with the critical exponent $\eta_T=1/3$ for reference. While (I) decays algebraically with an exponent smaller than 1/3, (II-III) decay exponentially. **c** Analogously, the orientational correlation functions $G_6(r)$ as a function of distance $r/r_0$ (solid orange), envelope fit (dashed) and fitted points (dots). The dotted black line represents the power-law with the critical exponent $\eta_6=1/4$. While (I-II) exhibit power-law behavior with $\eta_6$ smaller than 1/4, (III) yields a faster decay.

We stabilize a skyrmion lattice in a hexagonal geometric confinement of 100 µm edge length in a Ta(5)/Co$_{20}$Fe$_{60}$B$_{20}$(0.9)/Ta(0.07)/MgO(2)/Ta(5) stack with layer thickness given in nm in parentheses. We observe the skyrmions in real-time and -space with Kerr microscopy[7]. In Fig. 1a, we show Kerr microscopy snapshots (I-III) at different magnetic out-of-plane (OOP) fields, which controls the skyrmion size[17,19,20] and thus packing fraction.

We determine the local order parameter $\psi_6$ (as insets in Fig. 1a) and calculate the translational (Fig. 1b) and orientational (Fig. 1c) correlation functions to analyze the ordering. We fit the correlation functions $G_T(r)$ and $G_6(r)$ with a power-law decay, yielding exponents $\eta_T$ and $\eta_6$, respectively. Furthermore, we perform an exponential fit to determine the corresponding correlation lengths $\xi_T$ and $\xi_6$ (see Methods for details). Despite our finite system, we use the terms *solid*, *hexatic* and *liquid* measured in analogy to KTHNY theory to distinguish between different regimes.

For the densely packed skyrmions in snapshot I, the translational correlation $G_T(r)$ decays algebraically with an exponent $\eta_T$ smaller than 1/3. The orientational correlation $G_6(r)$ is almost constant. Hence, the system exhibits translational and orientational order which we identify as *solid*. For snapshot II, $G_T(r)$ decays exponentially while $G_6(r)$ is still decaying algebraically with an exponent $\eta_6$ smaller than 1/4. Thus, translational order has vanished while orientational order persists, which we identify as *hexatic*. In snapshot III, both $G_T(r)$ and $G_6(r)$ decay faster than with the critical exponent; therefore, we classify this unordered state as *liquid*.

Finite size effects occurring for the correlation functions, such as deviations for larger distances and the not clearly exponential behavior for snapshot III in Fig. 1c are described in the Supplementary Information (section S1 and Fig. S1). Furthermore, we see in Thiele model simulations that hexagonal confinements indeed stabilize ordering, particularly for smaller systems and with commensurability (Supplementary Information section S2 with Fig. S2). Regarding the qualitative processes and the critical exponents however, neither simulations nor experiments appear to be affected significantly by finite size effects. Therefore, we use the confinement for stabilization of the lattice while preventing local pinning centers from hampering lattice order macroscopically[15,16,33]. Thereby, we manage to reveal intrinsic effects and dynamics during a two-step melting expected from KTHNY theory.

In contrast to other systems in which 2D phases have been observed[14,15,29], our system is highly flexible in the sense that we can tune the skyrmions' size[17,38] and mobility[18] on the fly by the OOP magnetic field, in addition to the thermal Brownian-like diffusion[7]. Together with direct real-time observation in Kerr microscopy, this allows for a full real-time and -space analysis of the details associated with 2D phase transitions and critical phenomena, which have not been revealed in other systems. In the following, we perform the time-resolved quantitative analysis of the observed melting.

# Real-Time Quantification of the Melting

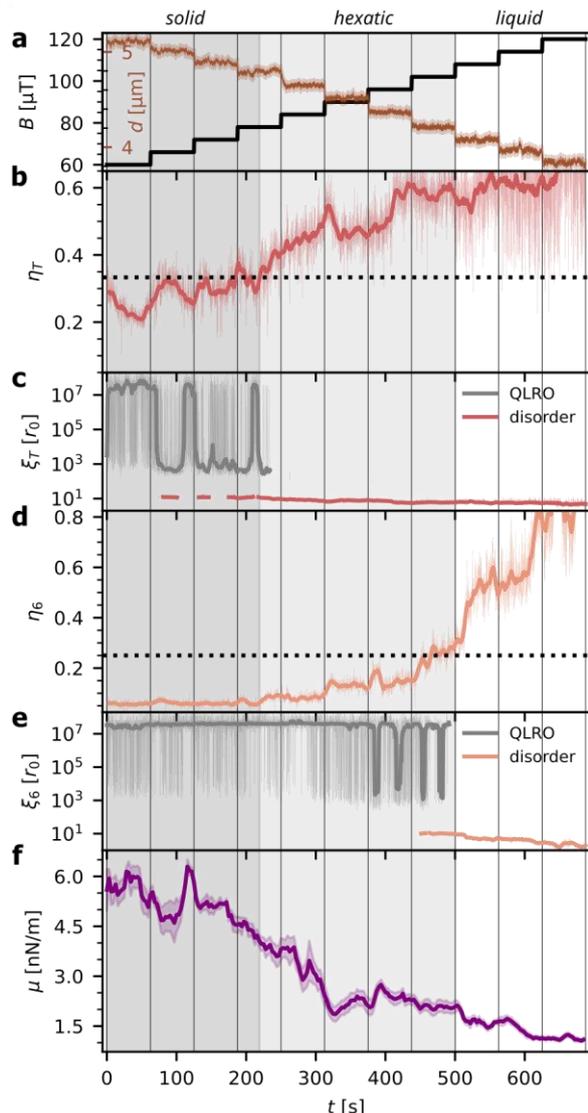

Fig. 2. Time-Resolved Quantification of the Melting. a The applied magnetic field B (black) is increased stepwise, directly causing the skyrmion diameter $d$ (brown, average and standard deviation) to decrease. Vertical black lines delimit the intervals of constant OOP fields in all subplots. b Exponent $\eta_T$ of $G_T$ and rolling mean over 6.25 s. The dotted horizontal line marks the critical exponent of 1/3. c Correlation length $\xi_T$ of $G_T$ and rolling median over 6.25 s. $\xi_T$ is well beyond the system size in the critical (QLRO) regime (gray) and within the system size in the disordered regime (red). d Exponent $\eta_6$ of $G_6$ and rolling mean over 6.25 s. The dotted horizontal line marks the critical exponent of 1/4. e Correlation length $\xi_6$ of $G_6$ and rolling median over 6.25 s for QLRO (gray) and disordered (orange) regime. The dark-gray, light-gray and white background represents the solid, hexatic and liquid regime, respectively. f Shear modulus $\mu$ of the skyrmion lattice determined from local lattice deformations, fit and standard deviation of the fit.

To capture the full melting process, we increase the magnetic field offset from 60 µT to 120 µT in steps of 6 µT every 62.5 s to shrink the skyrmions and thus reduce their packing fraction. In Fig. 2a, we visualize the OOP field (black) and the resulting skyrmion diameter (brown). The black vertical lines represent the time of the field changes in all subfigures. Since the shrinking skyrmions have effectively more space available, the number of accessible microstates and thus the configurational entropy is increased, which we exploit to drive the system to disorder.

Fig. 2b shows the time evolution of $\eta_T$, the black dotted line marks its critical value of 1/3. We find that initially, the exponent is below the critical value, signaling translational QLRO, which we identify as the solid regime. After 70 s, the exponent starts fluctuating around the critical value. At around 225 s, after further shrinking steps, $\eta_T$ permanently exceeds 1/3. However, the power-law exponent is only well-defined in the critical, ordered phase delimited by the critical exponent ($\eta_T$=1/3), where the decay becomes solely exponential. Fig. 2c shows the corresponding correlation length $\xi_T$. Note that $\xi_T$ must be determined differently for the critical (QLRO) and the disordered regime as shown in Fig. 2c. In the solid regime, we accordingly observe $\xi_T$ being orders of magnitude beyond the system size – effectively corresponding to divergence as predicted in KTHNY theory[25,26], highlighting the quality of the power-law fit. Reaching the critical exponent, the correlation length drops below the system size. Hence, the solid character given by translational QLRO has vanished.

Analogously, we plot $\eta_6$ and the associated correlation length $\xi_6$ in Fig. 2d and e, respectively. We see that $\eta_6$ starts well below its critical value of 1/4 (black dots) in the solid, meaning $G_6$ is almost constant (in contrast to $G_T$). At the point where the translational order vanishes, however, the orientational order persists as $\eta_6$ still stays below its critical value of ¼, which we identify as hexatic. In the hexatic regime, the exponent $\eta_6$ grows significantly until reaching 1/4 eventually, denoting the transition to a liquid. It fluctuates around the critical value after around 450 s and permanently exceeds it after 490 s. We visualize the three determined regimes by gray shading in Fig. 2.

To show the robustness of our analysis, we go beyond the entropy-mediated melting due to the skyrmion shrinking and exploit the tunability of our system further to provide a second, independent approach to drive the system across disordering transitions: By magnetic OOP field oscillations, we increase the diffusivity[18] (i.e., the effective temperature of the system) and thereby induce the melting. The results are consistent for the two approaches, details of a diffusivity-mediated hexatic-liquid melting is discussed in the Supplementary Information (section S3 and Fig. S3).

Finally, we can even probe the shear modulus $\mu$[39,40] analyzing local lattice deformations as characteristic behavior associated with the melting transitions is also exhibited by the elastic constants (Fig. 2f, see Methods for details). We see that $\mu$ is approximately constant in the solid. When translational QLRO vanishes, $\mu$ decreases significantly. However, $\mu$ is not vanishing since we must assume linear elasticity in our analysis, which becomes less applicable when reducing the packing fraction. Yet, qualitatively the resulting shear modulus clearly supports our conclusions.

Hence, we demonstrate that we can drive a finite lattice from solid order to a liquid via the hexatic regime and in particular, image the melting process directly with high time resolution. This allows us to access the dynamics of every skyrmion quasi-particle. With the full information of the individual skyrmion trajectories, we can uniquely identify the topological lattice defects and probe their dynamics, which is typically not accessible. The topological defects are the key feature of 2D melting as described in KTHNY theory. Therefore, we need to quantify the occurrence and dynamics of those lattice defects to fully understand the transitions.

# Dynamics of Topological Defects

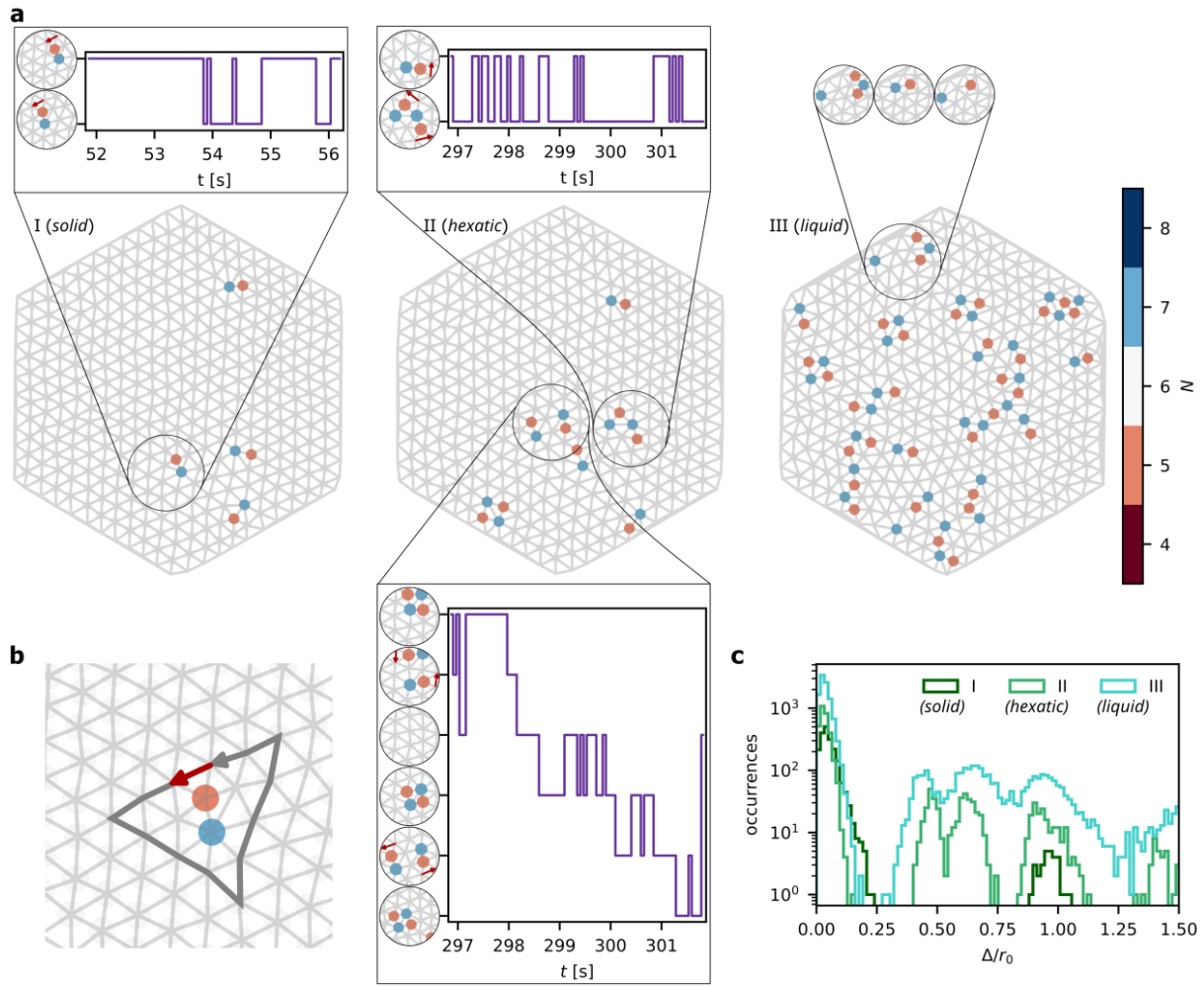

**Fig. 3. Fluctuations of Topological Defects. a** Lattice defects of snapshots I-III, colored by the number of lattice neighbors *N*. Snapshots I-III from Fig. 1 are used, which correspond to images of the melting shown in Fig. 2 at $t_I$=50.0 s, $t_{II}$=301.6 s and $t_{III}$=576.3 s, respectively. The gray net shows the nearest neighbor connections between the experimental skyrmions. The insets show how defect configurations can change between discrete states, the attached purple step plots visualize the fluctuations between those states in measurement time *t*. Exemplary Burger's vectors are drawn in red. **b** The four lattice vectors (gray) along each crystal direction would yield a closed path in a perfect lattice. However, to surround a dislocation counterclockwise, an additional lattice vector (red) is missing; it is identified as the Burger's vector. **c** Histogram of dislocation displacements $\Delta$ within 62.5 ms (1 frame) compared for snapshots I-III (using the statistics of 31 s).

The formation and dissociation of topological defects is the key feature of 2D melting in KTHNY theory. Our system allows us to observe the defect dynamics, gaining insight into the melting process in a unique way: The thermally activated diffusion of the skyrmions induces rich dynamics including lattice fluctuations and defect formations on sub-second time scales and dynamic imaging allows us to unambiguously identify the topological defects. Due to the time resolution of 62.5 ms, we can furthermore analyze the dynamics of the topological defects. Especially, we can tune the dynamics of both skyrmions and defects by the applied magnetic field. We note that previous observations of 2D phases lack sufficient resolution[14,31] or tunability of order and its fluctuations[15,29,30,34] to investigate the defect dynamics explicitly.

In Fig. 3a, we present the topological defect map for the previously defined snapshots I-III. In the dense-packed (solid) snapshot I, only four dislocations exist. We identify those dislocations as a consequence of the incommensurate number of skyrmions, combined with the underlying energy landscape of the sample[17]. These dislocations remain stable and may only hop between nearest-neighbor sites. An example of such a fluctuating dislocation is shown in the clipped inset connected to the black circle: The dislocation fluctuates back and forth between two discrete states. The discrete states are represented on the vertical axis in the black boxes in Fig. 3a, where the purple line shows the temporal fluctuation between the states in measurement time *t*. We see that the Burger's vector (red arrows in the insets) associated with the dislocation is preserved when the dislocation hops. An exemplary construction of the Burger's vector is drawn in Fig. 3b. In addition to the four dislocations present in the solid, dislocation pairs can form spontaneously and annihilate again with a short. The dislocation pairs are the characteristic clusters in the solid phase and have opposite Burger's vectors which cancel out to zero.

In the 5 s time-window before the hexatic snapshot II, new defects appear and exhibit more dynamics, also indicated in Fig. 3a: One inset shows how a dislocation can rearrange into two dislocations. Indeed, the total Burger's vector is conserved during the rearrangement. Furthermore, the second inset for snapshot II reveals how a dislocation pair can be created and annihilated, split up, merge and thereby move. Here, the local fluctuations result in at least the six different states as shown. Note that during the fluctuation, the total Burger's vector is always zero. However, the opposing vectors can change direction as the dislocation pair can split into either two horizontally or two vertically oriented dislocations. Again, the attached purple step plots depict the fluctuations between the distinct states over time for the respective insets.

In the disordered liquid snapshot III, the created and constantly rearranging defects form a complex pattern consisting of several defect clusters: The clusters still span dislocations and square dislocation pairs, but also isolated defects, linear defect chains and more complex configurations. The corresponding inset in Fig. 3a shows exemplarily how a dislocation can split into two disclinations or rearrange into an isolated defect plus a chain of three defects as a key feature of the liquid regime. Since dislocations and clusters in general can rearrange, split and merge in many ways and thereby increasingly interact with neighboring clusters, the number of observed states as well as their dynamics increases drastically. Eventually, the complexity does not allow us anymore to determine specific states and their fluctuations in time within a clipped region only.

This becomes more apparent when we investigate the dynamics of defect pairs: To this purpose, we match all defects into pairs of one *N*=5 and *N*=7 defect (see Methods for details). We consider nearest-neighbor matches as dislocations and link their occurrences to trajectories using *trackpy*[41]. Matches which are not nearest-neighbor connections are identified as disclinations. We provide a detailed visualization of the pair matching in the Supplementary Information (section S4 with Fig. S4), including a more generalized defect clustering in good agreement with previous simulation results[42]. Fig. 3c shows the histograms of the found dislocation displacements *Δ* within 62.5 ms for the snapshot I-III by using statistics of in total 31 s around each snapshot. We find that in the solid regime (snapshot I), the defects are stable and may only rarely hop by a nearest neighbor connection $r_0$. In the hexatic regime (snapshot II), nearest-neighbor hopping becomes more frequent and displacements different from $r_0$ occur. The inset of snapshot II in Fig. 3a may serve as illustration: The splitting of one dislocation into two causes a displacement of $r_0/2$ due to the associated reorientation, while dislocation pairs changing between two horizontal and two vertical pairs contribute with displacements of $\sqrt{r_0}/2$ each. In the liquid regime (snapshot III), an almost continuous distribution of displacements for the identified dislocations arises due to the complex rearrangements and interactions of defect clusters, signaling unbinding of topological defects.

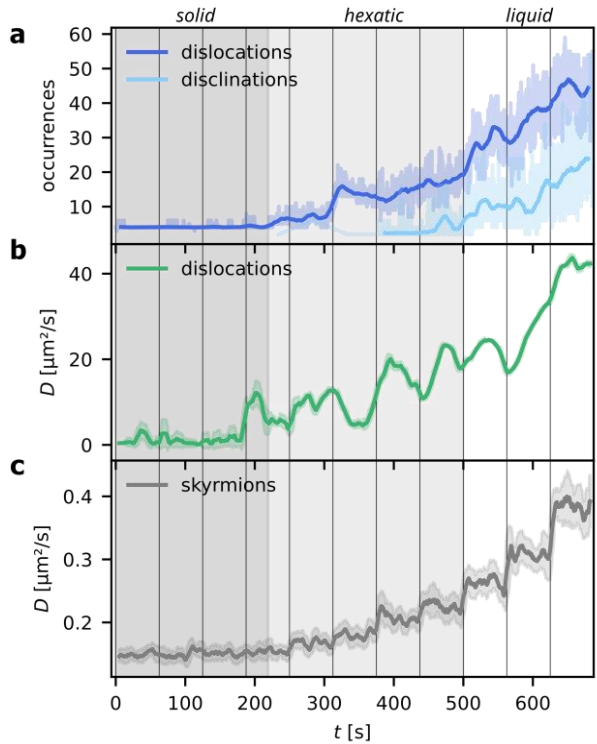

**Fig. 4 Time-Resolved Defect Dynamics a** Number of dislocations and disclinations observed during the melting. **b-c** The diffusion coefficient *D* of the dislocations (b) is up to two orders of magnitude larger than the one of the skyrmions (c). The dark-gray, light-gray and white background represents the predetermined solid, hexatic and liquid regime, respectively.

Our full real-time and -space analysis allows us to ascertain the quantitative evolution of dislocations and disclinations as shown in Fig. 4a. We find that the occurrence of dislocations (beyond the ones existing initially due to incommensurability) starts as predicted by theory at the transition from solid to hexatic. And occurrence of disclinations starts at the transition from the hexatic to liquid regime. Thus, our identification of dislocations and disclinations yields qualitatively good agreement with KTHNY theory.

Finally, the identification of dislocations as topological quasi-particles throughout the whole melting process allows us to investigate their dynamics. From the mean squared displacement (MSD) of the dislocations, we fit the diffusion coefficient *D* (see Methods for details) shown in Fig. 4b. We find that for the dislocations, *D* is almost constant in the solid regime and increases drastically when transitioning from solid to hexatic due to the dislocation unbinding. Within the hexatic regime, *D* continues increasing slightly as also fluctuations increase. When entering the liquid regime, *D* starts increasing drastically again due to the complex defect rearrangements. For comparison, we also determine the diffusion coefficient of the skyrmions (Fig. 4c). For the skyrmions, *D* increases continuously during the melting, does however not change its behavior qualitatively at the found transitions – highlighting the key role of the topological defects in mediating the melting. Instead, we see that the increasing skyrmion diffusivity enables the formation and rearrangement of topological defects, which drastically speeds up the dislocation dynamics: During the melting, *D* for the dislocations becomes up to two orders of magnitude larger than for the skyrmions forming the underlying lattice. This can be explained by the fact that dislocations are second-order (meta-) quasi-particles of the lattice (comprised of quasi-particle skyrmions) and can thus hop significantly even if skyrmions are only slightly displaced.

## Conclusion

In conclusion, we find the on-the-fly tunability of our system to be key to provide two unique and independent methods to controllably drive a 2D skyrmion lattice to disorder via a hexatic regime: both (i) by dynamically changing the skyrmion size (reducing the packing fraction) and (ii) by increasing the skyrmion diffusivity (i.e., effective temperature). We exploit a developed geometric confinement to locally stabilize the lattice and allowing us to capture the full intrinsic dynamics. Due to the extraordinary rich dynamics and flexibility, real-time and -space Kerr microscopy yields the necessary insights to understand the 2D melting process. We capture the emergence and dynamics of the characteristic topological defects mediating the phase transitions in KTHNY theory and quantify the dynamics of the involved dislocations. Thereby, we image the so far unrevealed key mechanism of 2D melting, including a meta-quasi-particle diffusion coefficient of the dislocations that is found to be orders of magnitude higher than the skyrmion diffusion. We quantify the correlation functions and the shear modulus, which behave in excellent agreement with KTHNY theory. Our work provides thus a new approach to access the key dynamics in 2D melting and opens up new possibilities to study intrinsic phase behavior, boundary effects and identify the role of topological defects in 2D systems.

# Methods

### Magnetic Multilayer Stack

The Ta(5)/Co$_{20}$Fe$_{60}$B$_{20}$(0.9)/Ta(0.07)/MgO(2)/Ta(5) multilayer stack (layer thickness in nm in parentheses with an accuracy better than 0.01 nm) is deposited by DC/RF magnetron sputtering using a *Singulus Rotaris* machine with a base pressure of 3×10$^{-8}$ mbar. The hexagonal geometric confinement is patterned by electron beam lithography (EBL) followed by Argon ion etching.

Interfacial Dzyaloshinskii-Moriya interaction (DMI)[43,44] is induced mainly at the Ta/Co$_{20}$Fe$_{60}$B$_{20}$ interface, the Co$_{20}$Fe$_{60}$B$_{20}$/MgO interface causes perpendicular magnetic anisotropy (PMA). The Ta(0.08) dusting layer is used to balance PMA and DMI[7,45] to host skyrmions, but also to optimize the energy landscape for skyrmion lattice formation and dynamics. We provide the OOP hysteresis loop in Fig. S5 of the Supplementary Information (section S5). The non-trivial topology of the observed bubbles is experimentally confirmed by spin-orbit-torque-driven skyrmion motion and supported by micromagnetic simulations[7,17,38].

### Skyrmion Stabilization and Imaging

A commercially available *evico magnetics GmbH* Kerr microscope is used to establish magnetic contrast with a resolution of 300 nm in space and 62.5 ms in time, using a blue LED light source and a CCD camera with a field of view of 200×150 µm$^2$. Magnetic fields can be applied both in in-plane (IP) and out-of-plane (OOP) direction. The alignment of the coils is optimized by aligning the shift of OOP hysteresis loops with and without IP field. The OOP magnet is custom-made to allow field control with sub-µT precision. The sample itself is placed onto a Peltier element directly on top of the coil for temperature control. The temperature is kept constant at 333.5 K and monitored by a Pt100 sensor directly next to the sample to ensures temperature stability better than 0.1 K.

Skyrmions are nucleated by applying an IP field pulse, which saturates the sample in IP direction at a constant OOP field. The resulting skyrmion lattice is equilibrated by an oscillating OOP magnetic field at 100 Hz with amplitudes up to 60 µT in addition to the constant OOP field offset.

We have direct and precise control over the skyrmion size via the applied OOP magnetic field[17,19,20]. The sizes of the individual skyrmions are detected by a machine-learning based pixel-wise classification[46]. Additionally, we can continuously tune the skyrmion diffusivity by sinusoidal OOP field oscillation in addition to the offset field[9,18]. In the presented melting procedures, we change the field conditions every 62.5 s (corresponding to 1000 frames) to obtain reasonable statistics.

Furthermore, the time interval is sufficient for the system to equilibrate while we can ensure stable measurement conditions during the whole melting protocol.

## Data Analysis

For the detection of the skyrmions from the gray-scale video and linking them to trajectories, we use the *trackpy* package[41] in *Python*. The obtained positions are used for every skyrmions to determine the local order parameter $\psi_6$ and its nearest neighbors applying a Voronoi tessellation, which automatically determines the lattice defects. Skyrmions at the edge of the system are neglected for the analysis of $\psi_6$ and lattice defects as their position at the edge produces artefacts in the Voronoi tessellation[47].

For all skyrmions, which are not located at the edge of the system, we determine a value for $G_T$ and $G_6$ with respect to all other skyrmions. We bin the values of the respective correlation and perform an average in every bin resulting in the distance-dependent correlation functions $G_T(r)$ and $G_6(r)$ as presented in Fig. 1a. The determination of the correlation function works for single frame images; however, we average the correlation functions of 10 consecutive frames (over 0.625 s) to reduce noise significantly. Therefore, all plots and fits of the correlation functions are performed on averaged data. To determine the decay of the translational correlation function, we fit $G_T(r)$ with a power-law decay $\propto (r/r_0)^{-\eta_T}$ as a function of distance $r$ in units of the skyrmion lattice constant $r_0$. We use the initial power-law fit to determine if the system is translationally ordered ($\eta_T$ below critical value of 1/3) or not ($\eta_T > 1/3$). In disorder however, the exponent $\eta_T$ is no longer well-defined since the decay of $G_T$ is now solely exponential. Therefore, we fit for the disordered cases the exponential $\propto exp(-r/\xi_T)$ instead of the power-law. Since the exponential term is technically also present in the ordered critical regime, we fit also for the occurrences of $\eta_T < 1/3$ an exponential, but as additional factor to the power-law. We use this additional factor in the fit as confirmation that the correlation length $\xi_T$ becomes infinite in the ordered regime. For the orientational correlation function $G_6$ we proceed analogously to determine the exponent $\eta_6$ as well as the correlation length $\xi_6$. However, the orientational correlation has a different critical value of $\eta_6=1/4$, which we use to determine whether the system is orientationally ordered ($\eta_6<1/4$) or not ($\eta_6=1/4$); and whether we fit the exponential as additional factor to or instead of the power-law, respectively.

In our system, we lack the possibility to apply stress forces to measure the elastic moduli directly. Instead, we analyze the local deformations of the lattice in real-space to estimate the shear modulus $\mu$[39,40,48]. We use as reference lattice a central skyrmion with six perfectly arranged nearest neighbors at positions $\mathbf{X}_i^{\text{ref}}$ with average lattice spacing. To this reference, we fit a local deformation tensor $\bar{\bar{\delta}}$ for every skyrmion and its neighbors in the experimental lattice, such that the squared distance

$$d^2 = \sum_i \left| \left( \bar{\bar{\delta}} \cdot \mathbf{X}_i^{\text{ref}} \right) - \mathbf{X}_i^{\text{exp}} \right|^2$$

between experimental lattice positions $\mathbf{X}_i^{\text{exp}}$ and tweaked reference is minimized. To extract the shear component, we decompose $\bar{\bar{\delta}} = \bar{\bar{\epsilon}} + \overline{\overline{R_\alpha}}$ to a symmetric strain tensor $\bar{\bar{\epsilon}}$ and an anti-symmetric rotation $\overline{\overline{R_\alpha}}$ by an angle $\alpha$. The diagonal elements of $\bar{\bar{\epsilon}}$ describe the strain along *x* and *y*, whereas the off-diagonal element is the shear component. In case of linear elasticity, a shear deformation is associated with a shear energy $E_{\text{shear}} = \frac{1}{2}(2\epsilon_{xy})^2 V \cdot \mu$, where *V* denotes the volume over which the shearing takes place (area spanned by the nearest neighbors in our case) and $\mu$ is the shear modulus. Assuming a Boltzmann distribution $P(E) \propto \exp(-E/k_B T)$ of the shear energy at the temperature *T*, we fit $\mu$ as the slope of

$$\log(P(E)) = -\frac{E_{\text{shear}}}{k_B T} + const. = \mu \cdot \left[\frac{1}{2}(2\epsilon_{xy})^2 \frac{V}{k_B T}\right] + const.$$

when calculating a histogram over the square bracket as measure of the logarithm of the shear energy distribution. The procedure requires the assumption of linear elasticity, which becomes less applicable in a less dense system, especially in the liquid. Also, the distribution of shear energies associated with the determined deformations is not perfectly Boltzmann-like, as already observed for colloid systems[40]. Since the dependence is not perfectly linear, we perform a set of fits over different ranges and use the standard deviation as error of the mean value. The determined lattice deformations, shear components of the strain tensor and shear energy distributions are presented and discussed in section S6 and Fig. S6 of the Supplementary Information.

The determination of topological defects follows directly from the Voronoi tessellation used for calculating the local ordering. Every skyrmion with a number of nearest neighbors $N$ different from 6, which is not located at the edge of the system, is identified as lattice defects. Since defects in the solid and hexatic regime almost only occur pairwise, identifying those pairs as dislocations is trivial. However, transitioning to a liquid, complex clusters of defects evolve. The complex appearance, including the interactions between defect clusters, makes the identification of the formal connection between defects impossible. To analyze the further evolution of defects, we establish a simplified approach of identifying pairs of defects: To every 5-defect $i$, we assign exactly one 7-defect $j$ and take the distance between the defects as $d_{ij}$. To establish unique pair connections, we minimize the total square distance

$$d_{\text{tot}}^2 = \sum_{ij} d_{ij}^2$$

associated with all possible connections $ij$ using the Hungarian method[49]. We identify a determined defect pair as a dislocation if the corresponding $d_{ij}$ is a nearest neighbor connection; otherwise, we identify the two connected defects as two disclinations. To study the dislocation dynamics, we keep only the center of mass of all identified dislocations and link them to trajectories with *trackpy*[41].

To evaluate the diffusion constant of the skyrmions at different times of the measurement, we determine the mean squared displacement (MSD)

$$\text{MSD}(t) = \langle [\mathbf{r}(t) - \mathbf{r}(t_0)]^2 \rangle = 2dDt$$

by calculating the square distance of skyrmion position $\mathbf{r}$ at time $t$ relative to the position at the time of initial occurrence $t_0$ and take the average over all skyrmions. The MSD is further related to the dimensionality $d$ of the system (here: $d=2$) and the diffusion coefficient $D$ over time $t$ in the case of normal diffusion[7,18,38]. Since we furthermore want to determine $D$ at any time $t_0$ with reliable statistics, we consider all trajectories present in a 10 s time window around $t_0$ and use the time of first occurrence as $t_0$. We then fit the first 1 s of the resulting MSD to determine $D$. For the dislocations, we proceed analogously but use all trajectories occurring in a time window of 31 s around $t_0$ to fit $D$ for statistical reasons because there are significantly fewer dislocations than skyrmions.

# Acknowledgements

This work was funded by the Deutsche Forschungsgemeinschaft (DFG, German Research Foundation) - SPP 2137 (project #403502522), TRR 173/2 Spin+X (projects A01, A12 and B02). The authors acknowledge funding from TopDyn. This project has received funding from the European Research Council (ERC) under the European Union's Horizon 2020 research and innovation program (Grant No. 856538, project "3D MAGiC") and under the Marie Skłodowska-Curie grant agreements No. 860060 ("MagnEFi") and No. 101119608 ("TOPOCOM"). The authors gratefully acknowledge the computing time granted on the supercomputer MOGON II and III at Johannes Gutenberg University Mainz as part of NHR South-West. M.A.B. was supported by a doctoral scholarship of the Studienstiftung des deutschen Volkes. E.M.J. acknowledges the Alexander von Humboldt Postdoctoral Fellowship. A.S. and M.K acknowledge support from the Norwegian Research Council through Grant No. 262633, Center of Excellence on Quantum Spintronics (QuSpin).

# Conflict of Interest

The authors declare no conflict of interest.

# Author Contributions

R.G. performed the Kerr microscopy measurements and experimental data analysis. J.R., S.M.F. and M.A.B. conducted the MD simulations; R.G., J.R, S.M.F and M.A.B analyzed the simulation data. F.K., M.A.S. and E.M.J. optimized and fabricated the multilayer stack. R.G. prepared the manuscript with the help of J.R., M.A.B., S.K. and E.M.J.; A.S., P.V. and M.K. guided and supervised the work. All authors have commented on the manuscript.

# Data Availability

Data is available from the corresponding author upon reasonable request.

# Supplementary Information

## S1. Finite Size Effects in the Experiment

Since the experimentally observed lattice is a finite system, it is expected to exhibit finite size effects. In contrast, measures to classify phases in KTHNY are derived for infinite systems. Therefore, we investigate in the following how our results align with those measures.

Analyzing the behavior of $G_T$ and $G_6$ at larger distances (Fig. S1a and b, respectively; both for snapshots I-III), we see that the correlation functions can increase with distance due to the lattice stabilization by the boundary, especially for distances beyond ten nearest neighbor distances $r_0$. In contrast to Fig. 1, we use a semilogarithmic plot to highlight the larger distances. The effect is stronger for the orientational correlation since the confinement with its commensurate shape provides primarily orientational stabilization. We observe the boundary effect on $G_T$ starting at distances of approximately 12 $r_0$; for $G_6$, an influence is visible already at 7 $r_0$. We therefore use those distances as limits for the fit range when we determine the decay exponents $\eta_T$ and $\eta_6$, respectively.

According to KTHNY theory, the correlation functions are expected to change their functional behavior qualitatively between power-law and exponential at the critical point determined by the critical exponents. We therefore compare the reduced $\chi^2$ of pure power-law and pure exponential fits during the entire melting procedure. For each distance bin of the correlation functions, we use the mean and standard deviation from the performed average over 10 frames (0.625 s) for the fit. In Fig. S1c, we show that a power-law fit for $G_T$ is indeed more favorable in the predetermined solid regime (dark-gray shading). Entering the hexatic regime (light-gray shading), the exponential becomes more favorable. Hence, the critical value of 1/3 does not appear to be affected by finite size effects. At $t>450$ s, the power-law fit yields decreasing values $\chi^2$ and seems to become favorable again; however, this is only because of the small values of $G_T$ deep in disorder, which therefore become increasingly prone to noise and the present boundary effects. For $G_6$, $\chi^2$ of the power-law fit is always smaller than of the exponential, thus indicating better description by a power-law behavior for all times. However, due to the boundary effect seen in Fig. S1b, the fit region is strongly limited and yields maximum 7 points; especially, we lack values over larger distances, which are decisive to analyze the functional behavior. In contrast to the critical QLRO regime, we therefore do not find a clear preference for either a power-law or exponential behavior but our results remain inconclusive regarding the exact form. In the experiment presented in Fig. S3h however, we find that $G_6$ is indeed exponential as soon as the stabilization due to the boundary is overcome; there, with sufficient diffusion.

For the translational correlation, we furthermore notice an anisotropy of the crystal order. Fig. S1e shows the Fourier transform (FFT) of the Kerr microscopy snapshot I. We clearly see discrete peaks (even in higher order) in sixfold symmetry as expected for a hexagonal lattice. The first-order peaks determine the reciprocal hexagonal lattice vectors used for the calculation of $G_T$. The stripes through the center (in six-fold symmetry) are an artefact caused by the hexagonal shape of the confinement. As a feature of the finite nature of our system, $G_T$ can evolve differently depending on the choice of the lattice vector used. Fig. S1f provides a comparison for the usage of the three lattice vectors in $G_T$ resulting in differently fast decays. The anisotropy I only possible on finite scales and depends on the specific location of lattice defects.

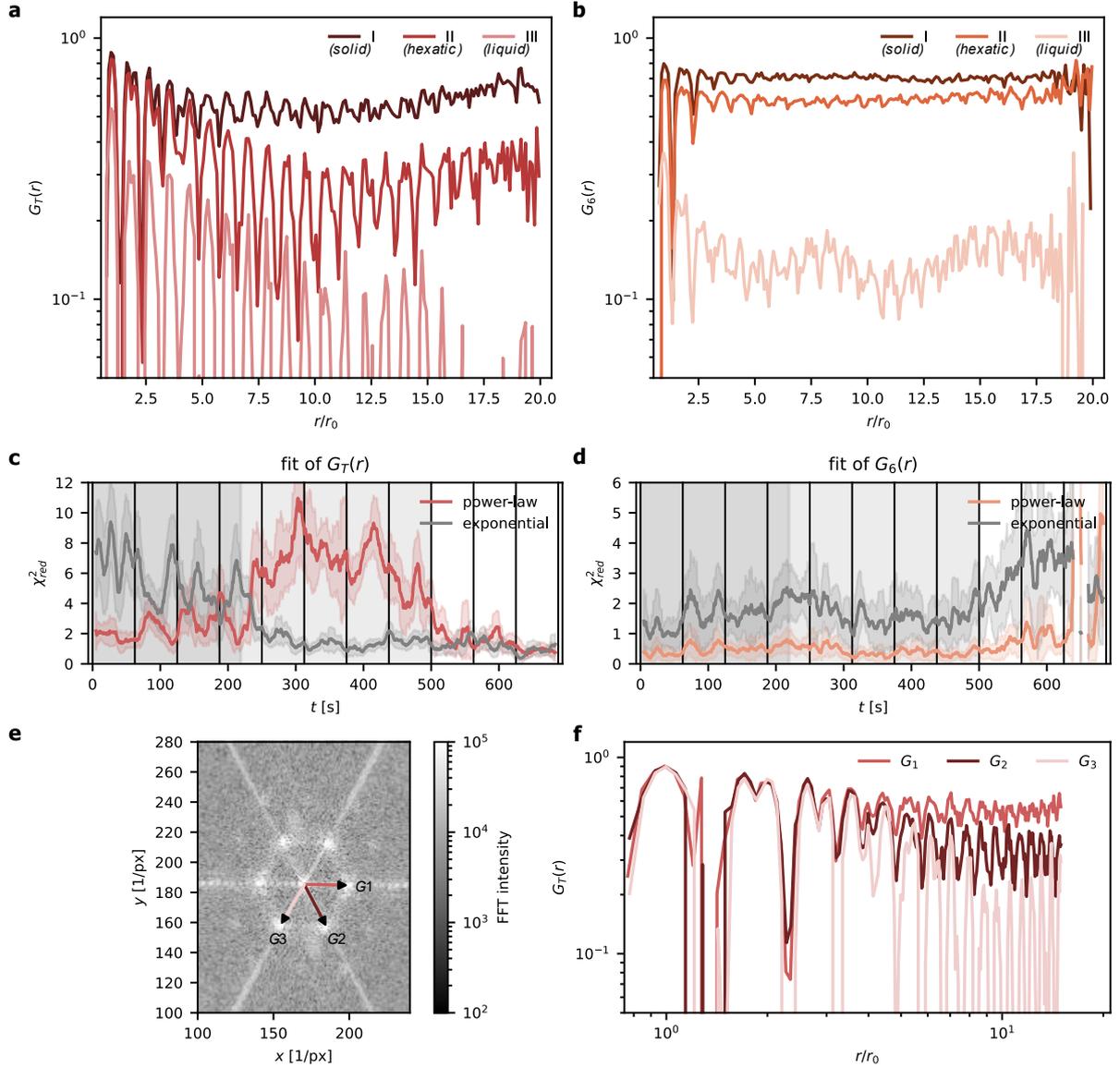

**Fig. S1. Analysis of Finite Size Effects in the Experiment. a-b** $G_T$ (a) and $G_6$ (b) for single-frame snapshots of snapshots I-III (as examples for solid, hexatic, liquid) as presented in Fig. 1, but up to the maximum system distance of ~$20r_0$. Note that in contrast to Fig. 1, we show a semilogarithmic plot to highlight the effect of larger distances: Due to the orientational stabilization by the hexagonal confinement, the correlation functions may increase again due to large distances being increasingly affected by boundary effects. **c-d** Reduced $\chi^2$ for fitting a pure power-law (colored) or exponential (gray) to the correlation functions $G_T$ (c) and $G_6$ (d) throughout the melting over time $t$. Average and standard deviation over 10 s are drawn. The dark-gray, light-gray and white shading represent the solid, hexatic and liquid phase, respectively. **e** Fourier transformed frame image of snapshot I exhibiting sixfold symmetry and determining the reciprocal lattice vectors $G_1$-$G_3$ (drawn as arrows). The six-fold lines are due to the hexagonal geometry of the system. **f** Due to the finite system size and the specific lattice defect configuration present, the translational correlation function may depend on the choice of lattice direction ($G_1$-$G_3$) as presented here for snapshot I.

## S2. Finite Size Effects in Simulations

To analyze finite size effects and their dependence on the system size, we perform computer simulations in the Thiele model[50] applied to magnetic skyrmions. In the case investigated in this article, the equation of motion reads[22] [Brems et al., *Phys. Rev. Lett.*, accepted (2024)]

$$-\gamma \mathbf{v} - G_{\text{rel}}\gamma \mathbf{e}_z \times \mathbf{v} + \mathbf{F}_{\text{therm}} + \mathbf{F}_{\text{SkSk}}(\{\mathbf{r}\}) + \mathbf{F}_{\text{SkBnd}}(\mathbf{r}) = 0$$

where $\mathbf{r}$ is the skyrmion position, $\mathbf{v}$ the skyrmion velocity and $\{\mathbf{r}\}$ indicates the set of all skyrmion positions. $\gamma$ indicates the damping (in the context of a Molecular Dynamics simulation, not to be confused with the Gilbert damping) and $G_{\text{rel}}$ is the relative Magnus force amplitude which is calculated as tangent of the skyrmion Hall angle. In this article, $G_{\text{rel}}$ and thereby the Magnus force is set to zero as the Magnus force affects only dynamics of the system but not static properties such as ordering. $\gamma$ is set to 1 (in simulation units). $\mathbf{F}_{\text{therm}}$ is thermal Gaussian white noise satisfying the fluctuation-dissipation theorem at a temperature of $k_B T = 1$ (in simulation units). $\mathbf{F}_{\text{SkSk}}$ and $\mathbf{F}_{\text{SkBnd}}$ indicate skyrmion-skyrmion and skyrmion-boundary repulsion. The skyrmion-skyrmion interaction is approximated by a $V(r)=r^{-8}$ potential with a cutoff distance of 1.8 simulation units[21] and the skyrmion-boundary interaction is using a fully repulsive Lennard-Jones potential

$$V_{\text{LJ}}(r) = 4\varepsilon \left[ \left(\frac{\sigma}{r}\right)^{12} - \left(\frac{\sigma}{r}\right)^{6} + \frac{1}{4} \right]$$

with $r_{\text{cut}}=2^{1/6}$ and $\varepsilon=\sigma=1$. Distances and the potential are not directly mapped to experimental values but are chosen to qualitatively reflect the behavior of skyrmions. The exponent of 8 is chosen to describe a similar potential steepness as found for skyrmions in a similar sample[22]. In contrast to more realistic ways of describing these interactions[22,23], the $r^{-8}$ potential features well-known liquid-hexatic and hexatic-solid phase transitions in continuum[21]. Densities $\rho$ in this simplified model are defined as skyrmions per unit length squared in simulation units. The equations of motion were integrated using an Euler algorithm

$$\mathbf{r}(t + \Delta t) = \mathbf{r}(t) + \mathbf{v}(t)\Delta t$$

with a time step of $\Delta t=10^{-4}$ implemented in the *HOOMD-blue* software package[51]. The system is initialized with either perfect hexagonal order (for commensurate numbers) or hexagonal order with a few particles missing (for non-commensurate numbers) and equilibrated for $10^6$ steps before running for $10^7$ steps with the trajectory saved every $10^4$ steps. Correlation functions are calculated and fitted individually for every saved step. Simulations of non-commensurate numbers are all averaged over at least 10 independent runs with independent initialization. Results of these simulations are presented as the mean over each individual simulation. Commensurate simulations use only one simulation run (except for 11 and 12 skyrmions per edge where we perform 10 runs) as there is only one specific starting condition and for large number a significantly longer computation time is required.

The main approximations of the Thiele model include that skyrmions are described as perfectly circular and of constant size as well as neglecting skyrmion creation and annihilation. These approximations are justified by the small amount of size polydispersity (standard deviation of the diameter: 11 %) as well as deformations found in the experiments. Also, the system investigated does not feature spontaneous skyrmion creation; skyrmion annihilation is only observed deep in the liquid regime. This makes the Thiele model applicable for a description of the skyrmions used in this article.

In Fig. S2a, we show for computer simulations of a hexagonal system with commensurate numbers of skyrmions (centered hexagonal numbers) that the hexagonal boundary condition stabilizes QLRO in the system. The stabilization becomes stronger for smaller systems, causing QLRO to persist down to lower skyrmion density. We plot the fraction of states observed in the solid, hexatic and liquid regime, respectively. For smaller systems, transition regions between the different regimes at identical parameters also become larger. For larger systems, the influence of the finite size effects decreases, the regimes occur separated and the densities where the system transitions between the regimes approach the values for an infinite system[21].

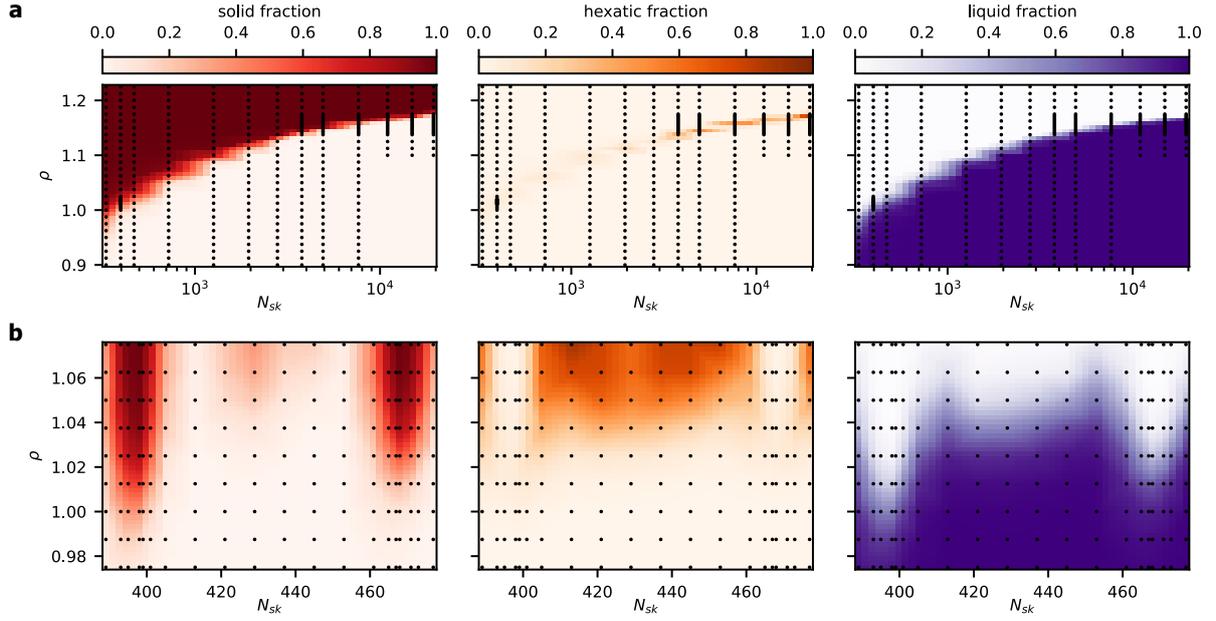

**Fig. S2. Ordering for different numbers of skyrmions in a hexagonal confinement. a** Ordering regimes in computer simulations of skyrmions in the Thiele model for different particle densities $\rho$ and system sizes using commensurate numbers of skyrmions $N_{sk}$ (centered hexagonal numbers) between 331 (10 skyrmions per edge) and 19441 particles (80 per edge). The black dots represent the parameters at which simulations are performed. The coloring of the parameter space denotes which fractions of the recorded simulations is in the respective regime; for visualization purposes, we show the linear interpolation of the determined fraction over the entire parameter space. The three regimes (solid, hexatic, liquid) exist for all system sizes and QLRO is stabilized by the hexagonal confinement: The smaller the system, the lower the density at which QLRO can still be stabilized; transition regions between different regimes also become larger. The square artefact between simulation parameters are due to the linear interpolation displayed on a logarithmic scale. **b** Analogously, ordering regimes of non-commensurate numbers of skyrmions. Between the commensurate states containing 397 (11 per edge) and 469 skyrmions (12 per edge), translational QLRO and thus the solid regime is strongly suppressed. Instead, the liquid and especially the hexatic regime are enhanced. In between, we find a half-commensurate state (alternating combination of 11 and 12 skyrmions per edge), QLRO is weakly stabilized.

In addition to the size scaling of the system, we also investigate how the commensurability of the particle number with respect to the confinement geometry[37] affects the ordering. Fig. S2b depicts similar simulations for various numbers of skyrmions around and between the commensurate (centered hexagonal) numbers of 397 (11 per edge) and 469 (12 per edge). As incommensurability enforces the existence of dislocations, it strongly suppresses translational QLRO, i.e. the solid regime; in return, the liquid and especially the hexatic regime are widened and enhanced due to quenched disorder[35,36]. Interestingly, we find a half-commensurate state alternatingly combining 11 and 12 skyrmions per edge, weakly stabilizing QLRO.

The experiments presented in the main text are performed with 401 skyrmions, i.e. slightly off commensurability. Indeed, the system exhibits the three different, well-separated regimes of order and is thus in good agreement with the simulations and KTHNY theory in general. In particular, we observe a relatively wide parameter space for the hexatic regime due to slight incommensurability and low pinning, while still benefitting from sufficient local stabilization due to the geometric confinement.

## S3. Lattice Melting by Increasing Diffusivity

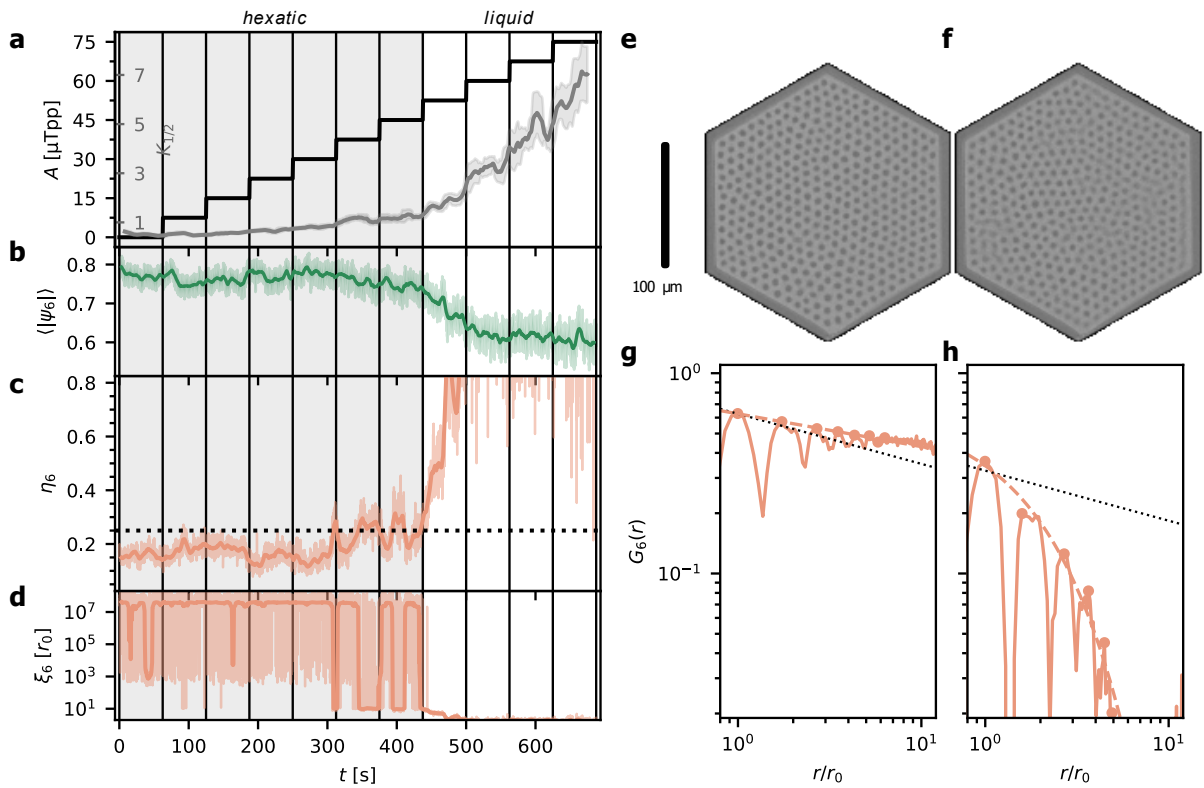

**Fig. S3. Lattice Melting by Increasing Diffusivity. a** The amplitude $A$ (peak-to-peak at 100 Hz) of the oscillating applied field is increased in steps (black) while the offset field is kept constant at 102 µT. This causes the diffusion coefficient $D$ of the skyrmions to increase (gray, rolling mean and standard deviation). bars represent the field intervals in all sublots. **b** Average absolute value $\langle|\psi_6|\rangle$ of the local order parameter, for each frame (shaded) and rolling average over 6.25 s (solid line). **c** Exponent $\eta_6$ of $G_6$ and rolling mean over 6.25 s. The dotted horizontal line marks the critical exponent of 1/4. **d** Correlation length $\xi_6$ of $G_6$ and rolling median over 6.25 s frames. **e-f** Kerr images averaged over 5 s at (e) $t$=75 s and (f) $t$=575 s. The smeared-out contrast visualizes the increased diffusivity in (f). **g-h** Orientational correlation functions $G_6$ (e) and (f), respectively. The black dots represent a power-law decay with the critical exponent of 1/4. The dashed line is the envelope fit and the dots are the points used for the fit. While in the hexatic case in (g), $G_6$ decays algebraically, it is now exponential in the liquid case shown in (h).

As a second approach to melt a skyrmion lattice, we increase the diffusivity of the skyrmions akin to an increase in thermal diffusivity mediated by an increasing temperature. We stabilize a similar lattice comprised of 378 skyrmions in the hexatic regime and keep the OOP offset field constant at 102 µT. In addition, we apply an OOP field oscillation at 100 Hz and increase the amplitude in steps in order to tune the skyrmion diffusivity[18] as shown in Fig. S3a. The corresponding decrease of the local order parameter is depicted in Fig. S3b. We again calculate and fit the correlation functions with an exponent $\eta_6$ (Fig. S3c) and a correlation length $\xi_6$ (Fig. S3d) as previously. We find that tuning the diffusivity, we can drive the $\eta_6$ from well below to above the critical value of 1/4, indicating a transition from hexatic to liquid. At the same time, $\xi_6$ is again dropping to the finite system size. In Fig. S3e-f, we show Kerr images averaged over 5 s for the hexatic and liquid case, respectively. The increasing diffusivity is clearly visible by the contrast smearing out. Thus, due to the averaging – which corresponds to the long exposure typically needed in other experiments[14] –, important details of the melting dynamics are lost – which we could however analyze in our system in Figs. 2-4 and Fig. S3. Fig. S3g-h finally show the orientational correlation functions corresponding to the hexatic and liquid snapshots in Fig. S3e-f, respectively. We corroborate that in the hexatic, $G_6$ is decaying algebraically with $\eta_6<1/4$ (black dots for reference). In the liquid, $G_6$ now also decays exponentially as expected as the increased diffusion is sufficient to overcome the stabilization due to the hexagonal boundaries.

## S4. Defect Clustering and Defect Pair Matching

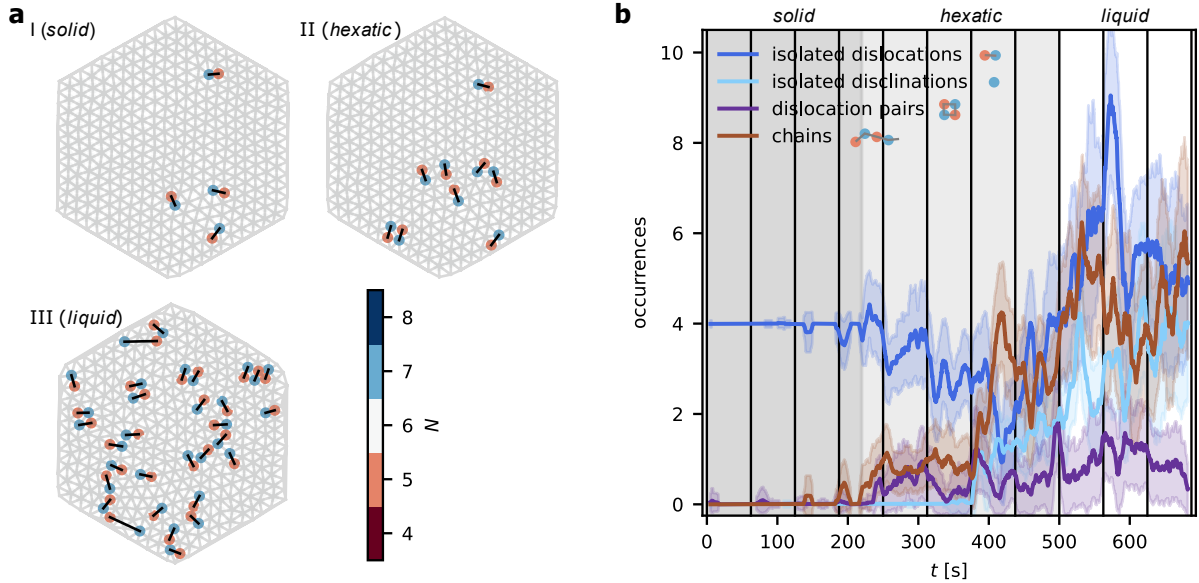

**Fig. S4. Defect Clustering and Pair Matching. a** Defect pair matching for snapshots I-III. The black lines represent the connection within a pair. While the matching is trivial in the solid and hexatic regime, the matching in the liquid is not obvious and obtained by distance minimization. **b** Occurrences of more specific clusters of defects occurring during the melting (rolling average and standard deviation over 10 s).

To detect and classify defect clusters, an approach based on the Voronoi tessellation of the system is used. For each particle with a neighbor number less (greater) than $N=6$, its neighbors are checked and neighboring particles with more (less) than $N=6$ neighbors are considered an associated pair. If no neighbor with an opposing neighbor number defect is found, it is classed as an isolated disclination. Otherwise, all particles within a graph of shared associations are considered part of the same cluster. Clusters consisting of exactly one 5- and one 7-defect are classed as isolated dislocations. Dislocation pairs are defined as consisting of exactly two 5- and two 7-defects where each defect is neighbor to both of the opposing defects. Chains are defined as any cluster consisting of at least 3 particles, which are connected linearly only.

Based on the Voronoi tessellation of the particle positions, particles that do not have exactly 6 neighbors are identified as defects and neighboring opposing defects are combined to clusters. This allows for the identification of different types of defect clusters to compare them to theoretical predictions[24–27]. While theory predicts only dislocation pairs in the solid phase, four stable 5-7-defect pairs (dislocations) are present. As the number of particles (401) is not commensurate with the confinement (closest match would be 397), at least one dislocation is needed to fit the additional particles. The system forms 4 dislocations which we attribute to pinning effects making certain defect configurations energetically favorable. In addition, we observe dislocation pairs as predicted. In the hexatic phase, dislocation pairs as well as chains proliferate, but defects are still bound in opposing pairs. The total number of defects again rises in the liquid phase and dislocations disassociate leading to isolated defects.

## S5. Hysteresis Loop

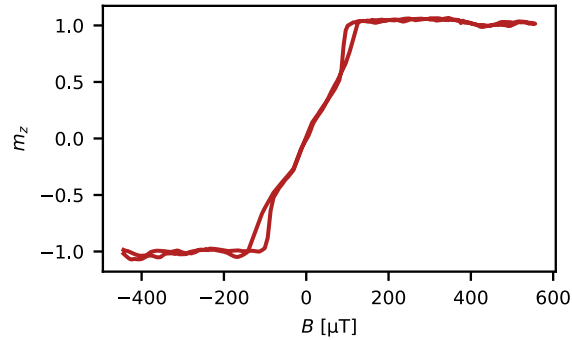

**Fig. S5. Hysteresis Loop.** The red line shows the MOKE intensity corresponding to the relative magnetization $m_z$ for a cycle of the OOP magnetic field $B$.

We measure the OOP hysteresis loop in terms of the relative OOP magnetization $m_z$ by the total intensity in Kerr microscopy over an OOP field ($B$) cycle. Note that the saturation field of the loop is below 200 µT, which is of the order of the earth magnetic field. We correct the surrounding magnetic offset field by shifting the hysteresis to be centered around zero. All field values given in the manuscript are corrected for the background field.

## S6. Shear Analysis

In Fig. S6a, we observe that during the melting, larger shear components of the strain tensor become possible and more likely. We furthermore present the determined local deformations for exemplary skyrmions in Fig. S6b for visualization. The deformation is fitted such that the deformation applied to the nearest neighbors of a model lattice (pink dots) matches the actual nearest neighbors (grey disks) best. From the local deformation over the area spanned by the nearest neighbors, we extract a shear energy contribution and its distribution (Fig. S6c) using statistics of 12.5 s. It becomes apparent that the distribution is not perfectly linear in the semilogarithmic plot. According deviations were found also in previous investigations[40]. We therefore fit the slope for varying low-energy and low-occurrence cutoffs and observe qualitatively consistent results with quantitative shifts. In Fig. 2f, we present the average value of the different fits and its standard deviation.

Furthermore, the shear modulus determined here does not vanish when translational QLRO vanishes. For the fit of the shear modulus, this would require a uniform distribution of shear energies in Fig. S6c. In this approach, the vanishing shear modulus is not captured we require linear elasticity in the formalism, which becomes less applicable during the melting. Also fitting the local deformations is more unstable and therefore prone to errors in the isotropic liquid. Yet, it is remarkable how well this very simplified method resembles a breakdown of the shear modulus.

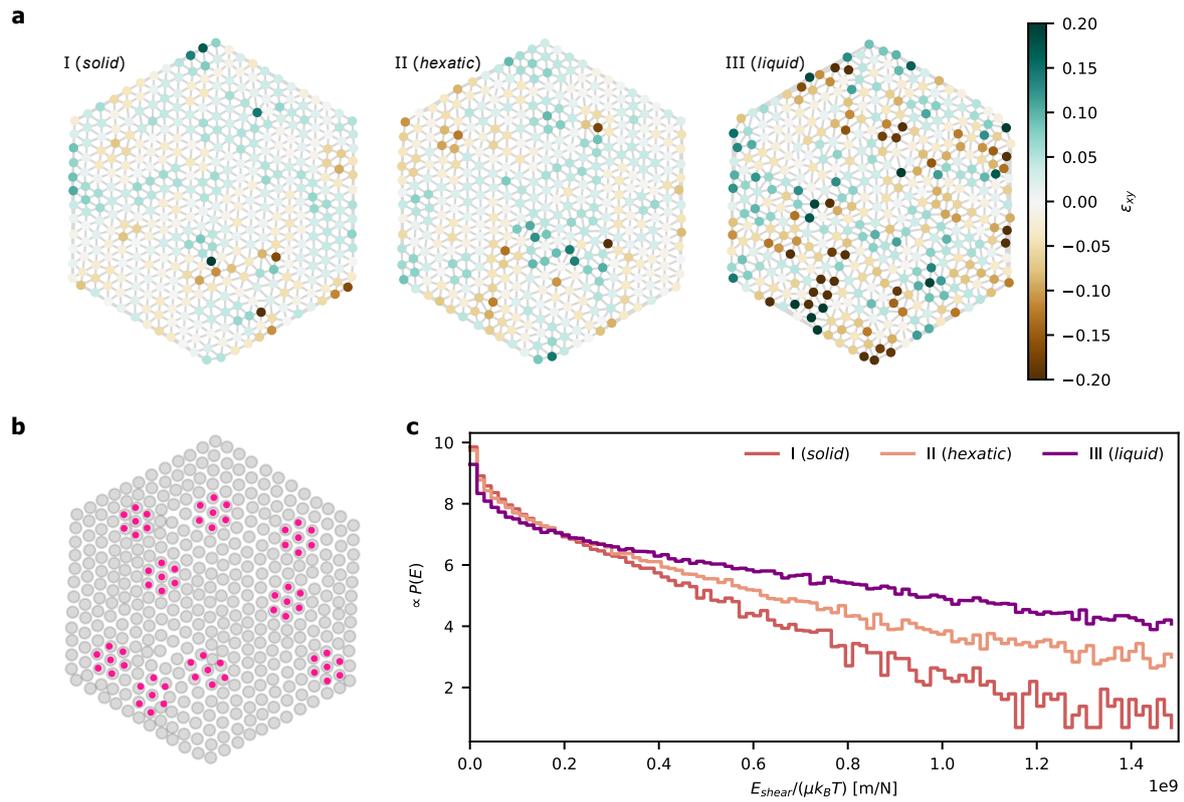

**Fig. S6. Shear Analysis of the Lattice. a** Shear components $\varepsilon_{xy}$ of the determined local lattice deformations for snapshots I-III. **b** Visualization of determined local lattice deformation at $t=0$. The gray circles denote the observed skyrmion positions. The pink dots represent the local deformation at exemplary sites: For every skyrmion position, a deformation of a local hexagonal lattice (up to the first nearest neighbors) is fitted to match the neighboring positions best. $\varepsilon_{xy}$ is extracted as the asymmetric component. **c** Every shear component is associated with a shear energy $E_{shear}$ in units of the shear modulus $\mu$ and $k_BT$, which is related to the Boltzmann energy distribution $P(E)$. The slope of the distribution determines the shear modulus $\mu$. The shear energy contributions are shown for the snapshots I-III using statistics over 12.5 s.